\documentclass[letterpaper, 10 pt, conference]{ieeeconf}  

\IEEEoverridecommandlockouts                              

\usepackage{graphicx}
\usepackage[width=.47\textwidth]{caption}
\usepackage{grffile} 
\usepackage{subcaption}
\usepackage{placeins}
\usepackage{array,multirow}
\usepackage{setspace,lipsum}

\usepackage{ifdraft}                                               
\usepackage{color}                                                 
\usepackage[normalem]{ulem}                                        
\definecolor{capri}{rgb}{0.0, 0.75, 1.0}
\definecolor{darkgreen}{RGB}{0,100,0}

\newcommand{\rg}[1]{{\scriptsize\color{blue}[RG: #1]}}

\renewcommand{\rg}[1]{}
\let\labelindent\relax 
\usepackage[inline]{enumitem}
\newcommand{\il}[1]{\begin{enumerate*}[label=(\roman*),mode=unboxed]#1\end{enumerate*}}

\newcommand{\fref}[1]{Fig.~\ref{#1}}       
\newcommand{\sref}[1]{\S\ref{#1}}          

\newcommand{\eg}[1]{\textit{e.g.,}~#1} %
\newcommand{\ie}[1]{\textit{i.e.,}~#1} %

\graphicspath{{Pictures/}} 

\usepackage{siunitx}
\DeclareSIUnit\feet{ft}
\DeclareSIUnit\week{w}

\usepackage{eurosym}

\makeatletter\let\@afterindenttrue\@afterindentfalse\makeatother

\renewcommand{\baselinestretch}{.97}

\title{\LARGE \bf
Wearable Embroidered Muscle Activity Sensing Device for the Human Upper Leg}

\author{R. B. Ribas Manero$^{1}$, J. Grewal, B. Michael, A. Shafti, K. Althoefer, Member, IEEE,\\ J. Ll. Ribas Fern\'{a}ndez$^{2}$ and M. J. Howard, Member, IEEE
\thanks{$^{1}$R. B. Ribas Manero, J. Grewal, B. Michael, A. Shafti, K. Althoefer and M. J. Howard are with the Centre for Robotics Research at King's College London, WC2R 2LS, London, UK - roger\_bernat.ribas\_manero, brendan.michael, ali.shafti, kaspar.althoefer, matthew.j.howard@kcl.ac.uk, $^{2}$ J. Ll. Ribas Fern\'{a}ndez is with the Human Anatomy and Embryology Unit at Universitat de Barcelona, Carrer de Casanova, 143, 08036 Barcelona, Spain - jribasfe@ub.edu}}

\begin{document}

\maketitle
\thispagestyle{empty}
\pagestyle{empty}

\begin{abstract}
Within the last decade, running has become one of the most popular physical
activities in the world. Although the benefits of running are numerous, there
is a risk of Running Related Injuries (RRI) of the lower extremities.
Electromyography (EMG) techniques have previously been used to  study causes of
RRIs, but the complexity of this technology limits its use to a laboratory
setting. As running is primarily an outdoors activity, this lack of technology
acts as a barrier to the study of RRIs in natural environments. This study
presents a minimally invasive wearable muscle sensing device consisting of
jogging leggings with embroidered surface EMG (sEMG) electrodes capable of
recording muscle activity data of the quadriceps group. To test the use of the
device, a proof of concept study consisting of $N=2$ runners performing a set
of \SI{5}{\kilo\meter} running trials is presented in which the effect of
running surfaces on muscle fatigue, a potential cause of RRIs, is evaluated.
Results show that muscle fatigue can be analysed from the sEMG data obtained
through the wearable device, and that running on soft surfaces (such as sand)
may increase the likelihood of suffering from RRIs.

\end{abstract}

\section{Introduction}\label{s:introduction}
Today, running is one of the most popular physical activities, as of Spring
2014 there were $65.58$ million runners in the USA accounting for a $20.53\%$
of the American population. Moreover, during the last decade, the number of
runners has grown considerably \cite{paluska2005overview}. The benefits of
running are numerous, ranging from increased life expectancy to preventing
chronic diseases or conditions (\eg{type-2 diabetes, cardiovascular disease,
and obesity}) \cite{o2010abc}. 

However, the practice of running also gives rise to Running Related Injuries
(RRI) of the lower extremities (\eg{aquiles tendinopathy, plantar fascitis, and
patellar tendinopathy}) \cite{van1992running}. The causes of RRIs are
heterogeneous, and changes in the biomechanical factors of the runner
(\eg{contact time, stride length}) have previously been investigated \cite{pinnington2005kinematic}. However, research is still lacking in
determining what role running surfaces have of changing these biomechanical
factors and hence causing RRIs. Particularly, how running surfaces influence
the fatigue levels of the leg muscles as a result of a running activity is not
clear. 

The most common method for estimating muscle fatigue, by examining the
electrical activity of the skeletal muscles (known as electromyography (EMG))
is however, difficult to perform in natural running settings. Examining EMG is
often done in a laboratory setting, requiring not only delicate expensive
equipment, but specialised anatomical knowledge for sensor placement, and
interpretation of results preventing non-specialists
from using EMG in other environments.

\begin{figure}[t!]\centering%
    \includegraphics[width=0.45\textwidth]{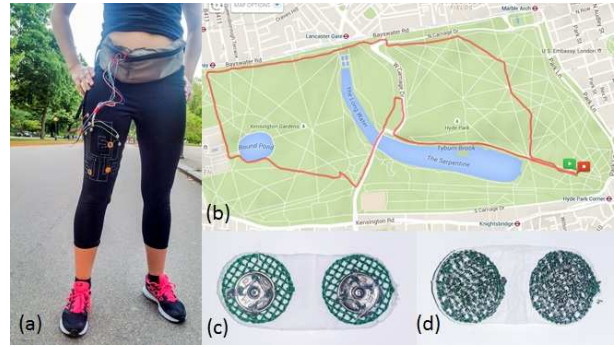}
	\caption{(a) participant wearing the jogging leggings with embroidered sEMG electrodes made for this study. (b) asphalt track route for participant 1. (c) and (d) top and bottom view of the embroidered sEMG electrodes respectively.}
    \label{f:intro}
\end{figure}

The objective of this study is to introduce a new wearable muscle sensing
device consisting of jogging leggings with embroidered surface EMG (sEMG)
electrodes capable of recording muscle activity data of the quadriceps group
(\ie{Vastus Medialis, Vastus Lateralis, and Rectus Femoris}) in a
non-laboratory setting. 
The wearable sEMG device allows \rg{collecting long term.../the collection of long term...} collection of long
term EMG data in a great variety of environments due to its compact design.
Furthermore, \rg{the use} use of the device does not require extensive prior anatomical
knowledge, as the sensing electrodes are embroidered into the garment at the
correct locations, following standards set by the SENIAM project
\cite{hermens1999european}. Thus, the user \rg{does not need to} need not be concerned about sensor
placement, and can simply pull the leggings on just like ordinary clothing.

Potential uses of EMG data collected in outdoor environments by the proposed
device, include understanding power demands for bipedal robots in outdoor
operation, as well as control of exoskeletons and prostheses outside of the
laboratory. In this paper, the device is demonstated as a platform for
monitoring muscle fatigue during endurance-based long distance
({\SI{5}{\kilo\meter}}) running tests on different surfaces (asphalt, sand and
an athletics track). Using data from the device, the dependency between running
surface and muscle fatigue is identified, whereby running on compliant surfaces
with greater damping (\eg{sand}) leads to the highest levels of muscle fatigue,
followed by very stiff surfaces with little damping features (\eg{asphalt}). These
results confirm expectations about likelihood of suffering from RRIs on these
different surfaces, and highlight the use of the proposed device for long term
observation of behaviour. \fref{f:intro} illustrates a participant wearing the leggings as well as a top and bottom up of the embroidered sEMG electrodes and the route followed during the asphalt trial by one of the participants.

\section{Background and Related Work}\label{s:related_work}
This section provides a brief overview of the functioning of electromyography,
and its use in measuring muscle fatigue. It also reviews related literature 
from laboratory based biomechanical studies on risk factors for RRIs.

\subsection{Methods for Electromyography}\label{ss:mus_emg}
Electromyography enables the force generated by the skeletal muscles to be
estimated by measuring their electrical stimulation from the central nervous
system (CNS) \cite{gonzalez2010emg}, via either invasive or non-invasive techniques. In regards to the former, \emph{intramuscular EMG} records activity by
inserting a fine wire into the muscle and measuring muscle signals. However,
while this results in a high quality signal its invasiveness makes it
unsuitable for many applications, especially in wearable sensing devices. 

A more suitable method is \emph{surface EMG} (sEMG) where electrodes
are placed on the skin surface above the target muscle. When following a
bipolar configuration, pairs of electrodes are placed on top of the belly of the muscle with a \SIrange{1}{2}{\centi\meter}
separation between the electrodes, and a reference electrode is placed on top of electrically neutral tissue.
The difference between the signals collected by each of the two electrodes
is amplified. 
Current commercial systems such as BTS Surface EMG (Vandrico, North 
Vancouver, Canada), and Tringo Wireless EMG (DelSys, massachusetts, USA) offer wireless EMG solutions. However, these present several disadvantages: 
\emph{(i)} they are expensive (ranging from \euro{448.00} to \euro{15,675.00}), \emph{(ii)} they are bulky (reducing their portability and affecting natural user motion), and \emph{(iii)} they rely on gel-based sEMG electrodes requiring anatomical knowledge for their placement and limiting their re-usability.

\subsection{Measuring Muscle Fatigue with EMG}\label{ss:mus_fati}
Due to the inherent issues with collecting sEMG data with standard sensors (\sref{ss:mus_emg}) the study of estimating and monitoring \emph{muscle fatigue} has been limited. Muscle fatigue is a consequence of muscle activity
\cite{gonzalez2012electromyographic}, to prevent cellular or organ failure
\cite{noakes2000physiological}.

Fatiguing of the muscles influences both the amplitude and the frequency
properties of EMG signals \cite{de1997use}, and during fatigue, further muscle
fibres need to be recruited (activated) in order to sustain the desired
performance. This is shown as an increment of the amplitude of the EMG signal.
Regarding the spectral features of the EMG signal, changes in the the firing
behaviour of the motor units and the shape of the Motor Unit Action Potential
(MUAP) result in a left-shifting of the power spectrum of the EMG signal. A
decrease of the instantaneous mean or mean frequency of the EMG signal
indicates the presence of muscle fatigue \cite{de1997use}.

A number of different approaches have been proposed in the literature in order
to elucidate muscle fatigue. These include 
\il{
	\item amplitude based parameters analysis, 
	\item spectral analysis, and 
	\item time-frequency analysis (\ie{wavelets ratios}).
} 
In amplitude based analysis,
the presence of muscle fatigue increases the instantaneous value of the EMG
signal with time \cite{gonzalez2012electromyographic}. Regarding the spectral
analysis, muscle fatigue results in a left-shifting of the power spectrum of
the EMG signal and hence a decrement of the instantaneous mean and median
frequencies is seen. Finally, for time-frequency analysis, the wavelets ratios
increase their value as the muscles get fatigued. Due to the isotonic nature of running, the collected sEMG signals cannot be considered as stationary and hence only amplitude based parameters and time-frequency techniques are suitable \cite{gonzalez2012electromyographic}.


\subsection{Muscle Fatigue as a Cause of RRIs}\label{ss:rris}
Among a number of other biomechanical factors, muscle fatigue has been shown to
significantly increase the likelihood of injury in dynamic behaviour such as
running \cite{apps2014individual}. To date, 
several studies have examined a number of factors in relation to this, including 
\il{
\item training characteristics (\eg{running volume}), 
\item anthropometric factors (\eg{foot strike}), 
\item biomechanical factors (\eg{contact time}), 
\item physiological variables (\eg{lactose level}), and
\item running-related medical history
}
\cite{saragiotto2014main}. However, largely due to the technological barrier of
taking measurement equipment out of the laboratory, few have examined these
factors in ordinary environments, as may be used by the general running
population \cite{tillman2002shoe}. This limits such studies to
simulating the running conditions, ignoring possible contributing factors.

One such factor may be the effect of the \emph{running surface} on fatigue, which is typically difficult to simulate in laboratory studies. Previous
research in this area has mainly focused on how these might affect the
biomechanical factors of the runner, using short ($<20\,m$) tracks \cite{ferris1998running,pinnington2005kinematic,tillman2002shoe}. As can be seen, despite the great amount of work in studying the biomechanics
of running, and the causes of musculoskeletal injuries, there is an apparent a
technological barrier in taking these studies into real world environments. In
the next section, the implementation of a new textile-based EMG system is
presented along with the signal processing techniques necessary for its use in
estimating muscle fatigue.

\section{Implementation}\label{s:methodology}
To overcome the limitations of collecting EMG data outside the laboratory
environment (as discussed in \sref{s:related_work}), this paper presents a new,
textile-based acquisition system. Specifically, the system consists of a pair
of jogging leggings with integrated sEMG electrodes, made from conductive yarns
directly embroidered into the fabric. 

\subsection{Design and Fabrication of the Jogging Leggings}
The leggings overcome many of the technical and user-specific limitations of
current sEMG monitoring devices. It includes built-in electrodes at appropriate
locations such that anatomical knowledge of optimal sensor locations is not
required when using the device.  It is also designed in mind of portability
for use in multiple environments, with built-in flexible low-weight printed
circuit boards (PCBs) to acquire raw sEMG signals in conjunction with an
Arduino micro-controller to digitise and store the collected data.

\subsubsection{Embroidered electrodes} The leggings make use of state-of-the-art
\emph{textile sEMG sensors}, recently developed in a parallel study at the Centre for Robotics Research (CORE) at King's College London, for measuring muscle activity in the legs.
As a precursor to this study, experiments were performed to elucidate the best fabrication process and design of these electrodes and characterise their behaviour for sEMG signal acquisition. Results show
that similar sEMG measurements can be obtained by using conductive textile
electrodes in place of gel-based disposable electrodes \cite{ali}. Comparisons
between the developed stainless-steal thread-based sEMG electrodes (using conductive thread Sparkfun DEV-11791,
\SI{3.28}{\ohm\per\meter}) and standard gel-based electrodes (Covidien Kendall
Arbo H124SG) show that, while there is greater noise presence when using
textile-based electrodes, muscle activity can still be observed and different
levels of muscle engagement distinguished. For this paper, this technology is
exploited to enable long term wearable measurement of  muscle activity through
an integrated textile device recording sEMG of the quadriceps muscles. 

\subsubsection{Circuitry and sensor placement}
\begin{figure}[t!]
    \centering
    \includegraphics[width=0.45\textwidth]{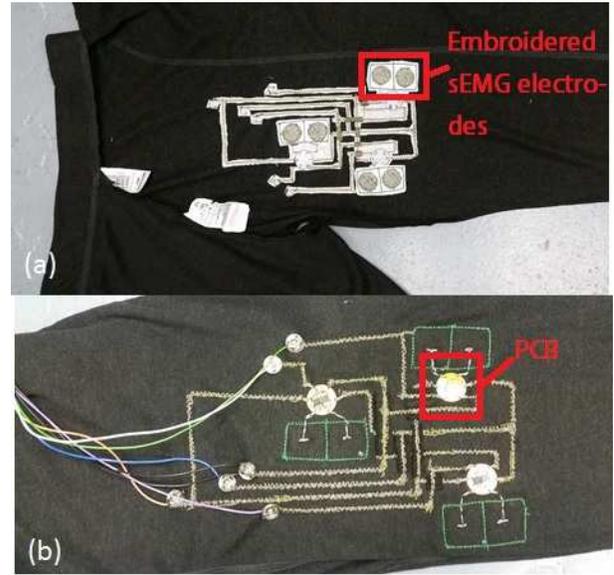} \label{fig:interior} 
	\caption{(a) the interior of the leggings. Notice how the connection lines were first sewn on top of stabilizer fabric. This type of fabric prevents the stretchy fabric of the leggings to stretch during the sewing process. (b) the exterior of the leggings. Note that the electric thread of the connection lines is sewn on top of the exterior face of the leggings in order to isolate them from the skin tissue.}\label{f:leggings}
\end{figure}

In the presented system, three pairs of electrodes are embroidered onto a pair
of jogging leggings using a Pfaff Creative 3.0 (Pfaff, Kaiserslautern, Germany)
sewing machine based on design variables obtained in \cite{ali} and according
to the indications of the SENIAM project following a bipolar configuration. The
placement area is tightened using extra thick felt fabric to enhance the contact
with the skin tissue. European medium (M) leggings size is selected and the in leg longitude of the participants' leg measured to assure the location of the sensors to be the one outlined by the SENIAM project. 
To prevent voltage loss due to the high resistance of the conductive thread,
the length of the connection lines is limited to \SI{20}{\centi\meter} for
which voltage losses are negligible. A double stitching pattern along with a
zig-zag pattern ensured robust and stretchable connection lines. The leggings
are powered at $\pm$\SI{3.75}{\volt} using a single rechargeable battery of
\SI{7.5}{\volt}. \fref{f:leggings} shows the final stages of the
fabrication process. Silicone sealant (102 RTV, Hylomar) is used to isolate
the connections of the power unit. Finally, a waterproof bag around the waist
is used to carry the batteries whilst running.


\subsubsection{Data acquisition}
\begin{figure}[h!]
    \centering
    \includegraphics[width=0.45\textwidth]{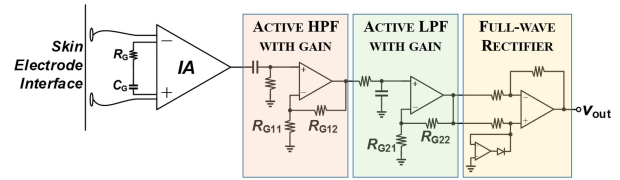}
    \caption{Circuit diagram of the sEMG acquisition PCBs. In red the active high pass filter, in green the active low pass filter and in yellow the full wave rectifier.}
    \label{f:schematics}
\end{figure}

Three custom PCBs were created to acquire the sEMG signal through each pair of
electrodes. Each PCB includes an instrumentation amplifier with adjustable gain
according to $G=1+(49.4k\Omega)/R_G$, a first order active high pass filter
(HPF), a first order active low pass filter (LPF), and a full-wave rectifier
connected in cascade. The instrumentation amplifier computes the difference
between the two collected signals and amplifies it. The cut-off frequencies of
the HPF and the LPF are \SI{20}{\hertz} and \SI{450}{\hertz} respectively. The
final PCB is flexible, with a thickness of \SI{1.65}{\milli\meter} and weight
of \SI{0.25}{\gram}. An Arduino micro-controller is used to sample the signal
from each PCB with a sampling frequency of \SI{1}{\kilo\hertz} and an SD shield
(Adafruit, NY, USA) is used to store the collected data in a text file with a
\texttt{.csv} format. \fref{f:schematics} shows the circuit diagram of the PCB.

\subsection{Signal Processing for Muscle Fatigue}\label{ss:sig_proc}
To quantify muscle fatigue from the sEMG data collected using the
sensor-embedded jogging leggings, the presented system includes a signal
analysis script written in MATLAB which implements the fatigue analysis methods
outlined in \sref{ss:mus_fati}. The script includes computations of the
\emph{instantaneous average rectified value} (iARV), \emph{the instantaneous mean average
value} (iMAV) and the \emph{instantaneous root mean square} (iRMS) value for
amplitude based analysis. For the spectral analysis approach, the instantaneous
mean and median frequencies, as well as the Dimitrov's spectral fatigue index
(FInsm5) \cite{dimitrov2006muscle} are implemented. For the time-frequency
analysis, the instantaneous WIRM1551, WIRM1M51, and WIRM1522 wavelets are
implemented. The wavelet forms to calculate the wavelet indices are chosen to
be the Symlet5 (sym5) and the Daubechies5 (db5) as according to Gozalez-Izal
\emph{et al.}\cite{gonzalez2010linear}. Due to the inherent characteristics of the running activity outlined in \sref{ss:mus_fati} and the fact that it has been extensively used in the literature \cite{gonzalez2012electromyographic}, the iARV signal was considered for assessing muscle fatigue. Such signal results of computing the \emph{average rectified value} (ARV) value for a \SI{0,2}{\second} window, where the ARV is computed as shown in (1).

\begin{equation}
\frac{1}{n} \sum_{n}^{} |x_n|
\end{equation}

Here, $x_n$ are the values of the sEMG signal and $n$ the number of samples.

\section{Experiments}\label{s:experiments}
This section presents experimental evaluations of the proposed device. A case study in which the device is used to detect muscle fatigue
in a long distance running experiment outdoors is introduced.




\subsection{Measuring Fatigue in Long Distance Running}\label{ss:fatigueexpt}
In this section, two evaluations are described. Firstly, an assessment of the
wearable sEMG monitoring jogging leggings outlined in  \sref{s:methodology} is
made, with an aim to investigate its suitability to natural environment motion
experiments. Secondly, the sEMG data collected is analysed using the custom
scripts in \sref{ss:sig_proc} to evaluate how the fatigue of the quadriceps
group varies depending on the running surface.

The experimental procedure is as follows. sEMG data is collected from $N=2$
participants (mean weight $75\pm$\SI{2.83}{\kilo\gram}, and mean height
$177\pm$\SI{5.66}{\centi\meter}).  For each participant, data is collected from
three running trials (each of \SI{5}{\kilo\meter} in length) on sand, asphalt,
and an athletics track, where participants run at their normal training speed.
Note, there is a minimum \SI{24}{\hour} resting period between trials, to allow
for muscles to recover.  In accordance with the methodology outlined in
\sref{s:methodology}, placement of the electrodes should not require anatomical
knowledge. As such, runners were instructed to wear the leggings, and the
electrodes were kept in the pre-built, standard position. 

Before beginning the experiments, participant data is collected, including
fitness level, age, height, weight, sex and running gear. A pre-screening test
is performed to ascertain whether the participants were eligible for the
experiments or not (PAR-Q\&U questionnaire approved by the Canadian Society for
Exercise Physiology). Ethical approval for the study was obtained prior to
experiments from the King's College London Research Ethics office \footnote{Ethics reference number LRU-14/15-1681}.

\begin{figure}[t!]
    \vspace{-0.5cm}
    \hspace{-0.25cm}
    \begin{tabular}{m{2.4cm}m{2.4cm}m{2.4cm}}    
     \raisebox{1.5cm}{\includegraphics[width=3cm,height=9cm]{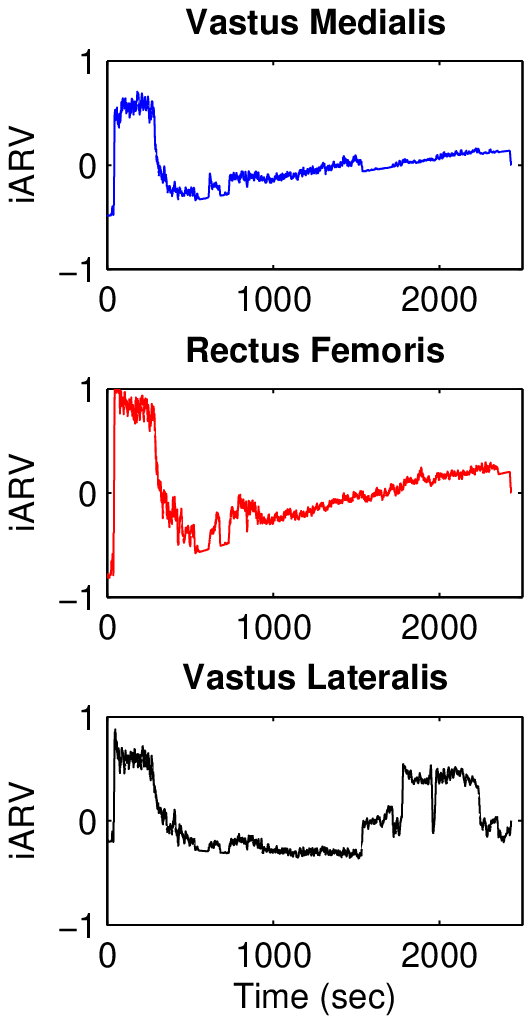}}\label{asphalt1} &
    \raisebox{1cm}{\includegraphics[width=2.9cm,height=9cm]{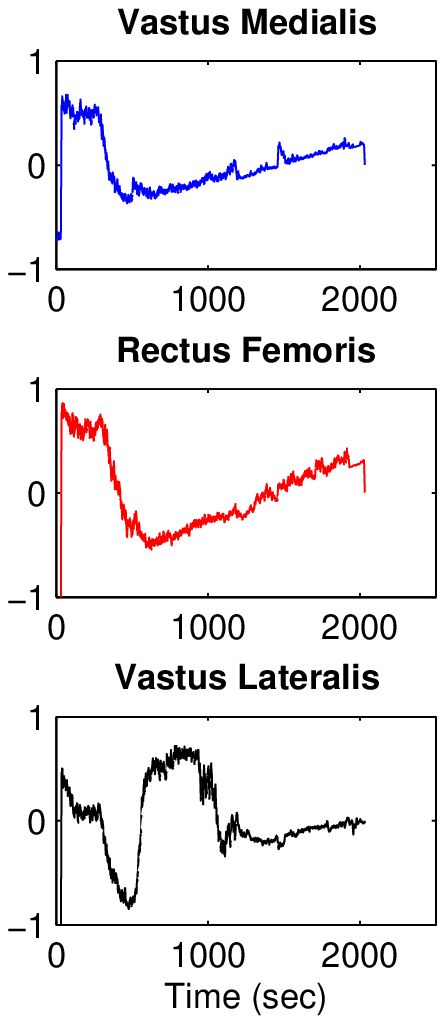}}  \label{asphalt1} &
    \raisebox{1cm}{\includegraphics[width=2.9cm,height=9cm]{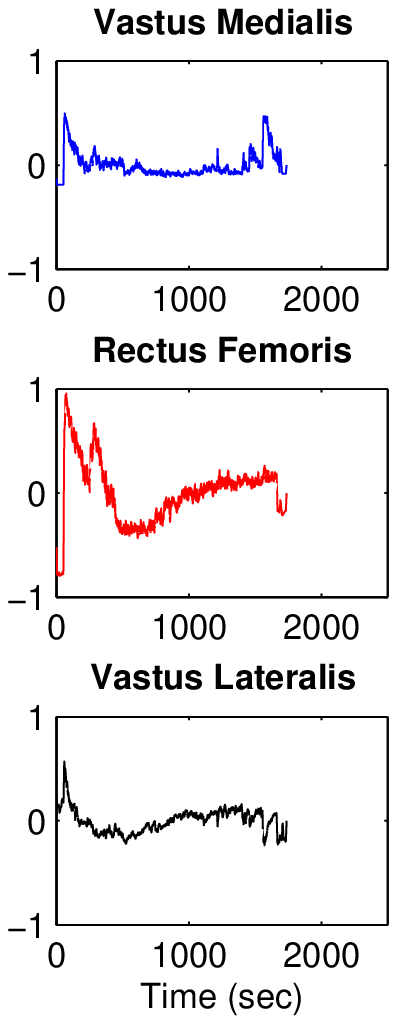}} \label{asphalt1}    
    \end{tabular}\vspace{-1.75cm}
	\caption{iARV of the quadriceps muscles whilst running on asphalt (blue), sand (red), and the athletics track (black). Note that the iARV signal increases with time due to the effect of muscle fatigue. 
    }\label{fig:fatigueEMG}
\end{figure}

\fref{fig:fatigueEMG} shows the iARV for participant 1 whilst running on the
three different surfaces. As explained in \sref{s:related_work}, the amplitude
of the sEMG signal increases in presence of muscle fatigue as more muscle
fibres are recruited. This phenomena is observed for the three surfaces with
varying rates of increase, as the iARV value of the sEMG signal increases with
time. Table 2 summarises the  percentage increase of the iARV signal for each
muscle whilst running on each surface between the beginning and end of the
running trials. It can be seen that running on the sand surface leads to the
maximum amplitude increment of the iARV signal for the three muscles whereas
the athletics track led to the minimum increase after the \SI{5}{\kilo\meter}
trials. Similar results are observed for participant 2, indicating that running
on sand increases the likelihood of suffering from RRIs when compared to
asphalt or athletics track surfaces. 

\begin{table}[t!]
\centering
\label{my-label}
\begin{tabular}{cccc}
\hline
Muscle           & Asphalt  & Sand     & Athletics Track \\\hline
Vastus Medialis  & 100.04\% & 127.71\% & 54.9\%          \\
Rectus Femoris   & 100.02\% & 126.75\% & 121.22\%        \\
Vastus Lateralis & 99.14\%  & 100.07\% & 35.9\%         
\end{tabular}
\caption{Percentage increase of the iARV signal for each muscle whilst running on asphalt, sand and the athletics track surfaces between the beginning and end of the running trials.}
\end{table}

\subsection{Further analysis of iARV}
As shown in \sref{ss:fatigueexpt}, wearable jogging leggings with in-built embroidered
electrodes are a practical and effective method for the collection of sEMG
signals in natural environments. However, due to the uncontrolled nature of these
environments, there are a number of factors that could confound the results of
the proceding section. This section discusses these issues, and their likely effect
on the results presented so far.

\subsubsection{Second wind effect}
Looking at the signals in \fref{fig:fatigueEMG}, it can be observed that
the iARV is greater at the beginning of all the trials and then decreases
progressively before increasing in presence of muscle fatigue. This can be
explained as a result of the second wind effect, which increases the serotonin
levels of the participant after the first minutes of running
\cite{gondola1982psychological}. This results in a decrement of the fatigue
levels of the muscles at the beginning of an exercise.

\subsubsection{Effect of Perspiration}
The second issue relates to the effect of perspiration on the data recorded.
During the course of experiments involving physical activity, the conductivity
the user's skin tissue varies linearly with the number of active sweating
glands \cite{thomas1957relationship}. In other words, the higher the sweating
levels of the participant are, the more conductive their skin will be. In the
device created for this study, the conductive thread wiring is not \rg{isolated} insulated
and hence the conductance of the wiring changes, altering the behaviour of the
circuit by increasing the inherent offset at the output of the PCBs. This
phenomena is observed for all the trials and is particularly noticeable for the
Vastus group muscles. Note that, the outdoor temperature for the first two trials (asphalt and sand)
on the day of collecting data was \SI{18}{\degreeCelsius} and
\SI{17}{\degreeCelsius} respectively whereas \SI{27}{\degreeCelsius} was
registered during the athletics track trial suggesting higher sweating levels
during the last trial and, therefore, a greater percentage increase of the iARV signal. This suggests that the athletics track might result in lower fatigue levels than the ones measured. In  spite of this phenomena, the classification of the surfaces remains the same as the athletics track still led to the lowest levels of muscle fatigue independently of the effect of the sweating.


\section{Conclusion}
In this study, the design of a new minimally invasive muscle sensing device
capable of recording muscle activity data in natural environments is
created. The device offers wearable capabilities that allow
long-term collection of data by inexperienced users. As explained in
\sref{s:introduction}, changes in biomechanical factors such as muscle fatigue
increase the likelihood of suffering from RRIs. A common way to study those
changes is through EMG technology. However, current technology limits studies of muscle EMG to laboratory setting and requires operation by users with anatomical knowledge. 

The experiments presented here show that wearable sEMG technology is feasible
and that it has practical research applications. The wearable sEMG device is cheap, robust,
easy to build and suitable for the every day user. As a proof of concept, $N=2$
runners took part in three \SI{5}{\kilo\meter} running trials on three
different surfaces (asphalt, sand and athletics track) and with sEMG data recorded using reusable fabric-based electrodes. This equipment is then used in an outdoors setting to elucidate muscle fatigue. The results showed that
different levels of muscle fatigue could be detected when running on different
surfaces. They matched the expectations that running on the athletics track
leads to the lowest levels of muscle fatigue, followed by the asphalt track and
finally by the sand track as a direct consequence of muscle activity as outlined in \cite{pinnington2005kinematic}. This, in turn, suggests that running on sand
increases the likelihood of suffering from RRIs when compared to the asphalt
and athletics track surfaces.

There are many possible additions to the design
which might improve performance. More robust PCBs without full wave
rectification would make the overall design more resistant and allow time-frequency
fatigue analysis techniques. Statex (PA66) and Loctite Hysol GR 9800 (Loctite,
Düsseldorf, Germany) could be used for encapsulation of the PCBs
\cite{linz2005embroidering}. Isolation of the conductive thread using Aleen's
flexible stretch fabric glue (Duncan Enterprises, California, USA) would make
the whole design robust against sweating.




\section*{Acknowledgment}
The authors would like to thank Karina Thompson (www.karinathompson.com) for
her help and support in digital embroidery. This work was supported by the UK
Crafts Council Parallel Practices project.


\bibliographystyle{IEEEtran}
\bibliography{IEEEabrv,Bibliography}

\end{document}